\documentclass[runningheads]{svmult}

\usepackage{makeidx}   % allows index generation
\usepackage{graphicx}  % standard LaTeX graphics tool
\usepackage{subeqnar}  % subnumbers individual equations
\usepackage{multicol}  % used for the two-column index
\usepackage{physprbb}  % modified textarea for proceedings,
\makeindex             % used for the subject index
\begin{document}
\title*{Search for sub-stellar companions using AO -\protect\newline first results obtained with NAOS-CONICA}
\toctitle{Search for sub-stellar companions \protect\newline using AO first results
obtained with NAOS-CONICA}
% allows explicit linebreak for the table of content
%
%
\titlerunning{Search for sub-stellar companions}
% allows abbreviation of title, if the full title is too long
% to fit in the running head
%
\author{M. Mugrauer\inst{1}
\and R. Neuh\"auser \inst{1} \and E. Guenther \inst{2} \and W. Brandner\inst{3} \and J. Alves
\inst{4} \and M. Ammler \inst{1}}
\authorrunning{Mugrauer et al.}
% if there are more than two authors,
% please abbreviate author list for running head
%
%
\institute{AIU Jena, Jena 07745, Germany \and TLS, Tautenburg 07778, Germany \and MPIA, Heidelberg
69117, Germany \and ESO, Garching 85748, Germany}

\maketitle              % typesets the title of the contribution

\begin{abstract}
We present first results of our NAOS-CONICA search for close substellar companions around young
nearby stars. This program was started only a few months ago. We have obtained 1$^{st}$ epoch
images of several targets which are unpublished young stars ($<$100\,Myrs), hence ideal targets to
look for planetary companions. For one target star we could even take a 2$^{nd}$ epoch image. By
comparing both images we could look for co-moving companions of the target star. Those data show
clearly that the detection of planetary companions (m$<$13\,$M_{Jup}$) inward a saturn-like orbit
(r$<$10\,AU) is feasible with NAOS-CONICA and in addition that the astrometric confirmation of
those companions is doable with only a few weeks of epoch difference.
\end{abstract}

\section{Speckle and AO imaging - the beginning}
We have taken deep, high dynamic range images of most stars in the TW Hydra, Horologium, Tucana and
$\beta$ Pic groups (age=10..40\,Gyrs, d=10..60\,pc). Such young nearby stars are well suited for
direct imaging of substellar companions, both brown dwarfs and massive planets (Neuh\"auser et al.,
2003[1]). The target stars are located in young star forming regions, hence companions are young
and therefore still self-luminous. Due to the proximity of the target stars a high spatial
resolution can be achieved. We started our search by using seeing limited imaging with SOFI and
speckle technique with the MPE speckle camera SHARP, both at the ESO 3.5m NTT on La Silla.
Companion candidates detected in 1$^{st}$ epoch images must be confirmed by 2$^{nd}$ epoch images
(proper motion) and follow-up spectra, in the latter case done with ISAAC at the VLT. From
non-detection and detection limits, we could conclude that massive planets (m$>$5\,$M_{Jup}$) in
wide orbits (a$>$50\,AU) are rare and appear around less than 9\,\% of the stars. Finally our
results have shown that the detection of extrasolar planets around young nearby stars was indeed
already possible without AO, i.e. with SHARP at the 3.5m\,NTT, but only at large separation outside
50\,AU.

\section{Imaging with NAOS-CONICA - first results}

Only a few months ago we have started a deep high dynamic range survey for very close planetary
companions using NAOS-CONICA (NACO) at the VLT. Our survey is additional to other direct search
programs because the targets in our program are unpublished (to be young) hence most likely not
observed by other groups already searching for planets. We started our observations (1$^{st}$
epoch) in May 2003 and continued it in Sep. 2003. Many stars were observed and several companion
candidates were detected only few arcsecs separated from the target stars.
\begin{figure}[hbt]
\begin{center}
\includegraphics[width=0.69\textwidth]{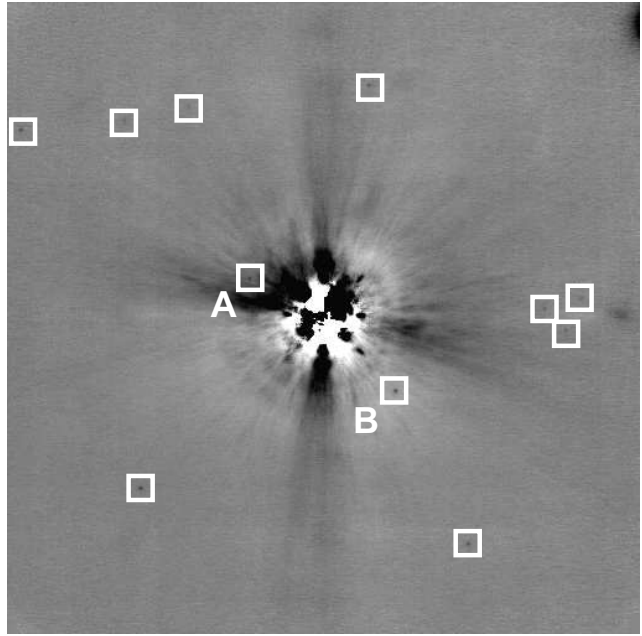}
\end{center}
\caption[]{We show here the 2$^{nd}$ epoch PSF subtracted NACO image of the young nearby star
HD\,180617. The FOV is 13.6x13.6\,arcsecs. The input seeing was about 0.9 arcsecs. Many faint
companion candidates are detected in K$_{S}$. They are marked with small white boxes. The closest
two objects (A, B) are also shown in the dynamic range plot, in Fig.2.} \label{eps2}
\end{figure}
We use a pixelscale of 13.25\,mas which gives us a good sampling of the VLT PSF and provide
concurrently the highest spatial resolution achievable with NACO. A short individual integration
time (0.347\,s) is used to avoid saturation as well as possible. That's why we have to superimpose
many of those short integrated frames to obtain deep infrared images. Fig.1 shows a NACO image of
HD\,180617, one target star which we could observe twice in 1$^{st}$ and in 2$^{nd}$ epoch. We have
subtracted the stellar PSF to measure astrometry and photometry of all detected faint companion
candidates. In Fig.2 (left) the measured dynamic range for this image is shown. If bound all
detected objects should have planetary masses (m$<$13\,$M_{Jup}$) assuming a conservative age of
100\,Myrs and using models of Baraffe et al. (2003)[2]. This analysis shows clearly that the
detection of planetary companions inside a saturn-orbit (10\,AU) is possible. After PSF subtraction
gaint planets down to jupiter-orbits (5\,AU) come into range. In the given epoch difference of 4
months we could measure the spatial motion of all detected objects including the bright primary.
The detected companion candidates are clearly not co-moving, hence they are not real companions of
HD\,180617. Nevertheless those data can be used to measure the proper and parallactic motion of
HD\,180617. Detected background stars are used as reference points for relative astrometry. The
parallax has a large stake in the entire motion of the star (see dashed and solid lines). The
stellar motion in the given time span is more than 100 times larger than the astrometric
uncertainties obtained with NACO. This illustrates well that few weeks of epoch difference are
enough to find faint co-moving companions, hence the detection of co-moving companions is possible
in only one observing period with NAOS-CONICA if the proper motion is large enough.

\begin{figure}[hbt]
\begin{center}
\includegraphics[height=0.39\textwidth]{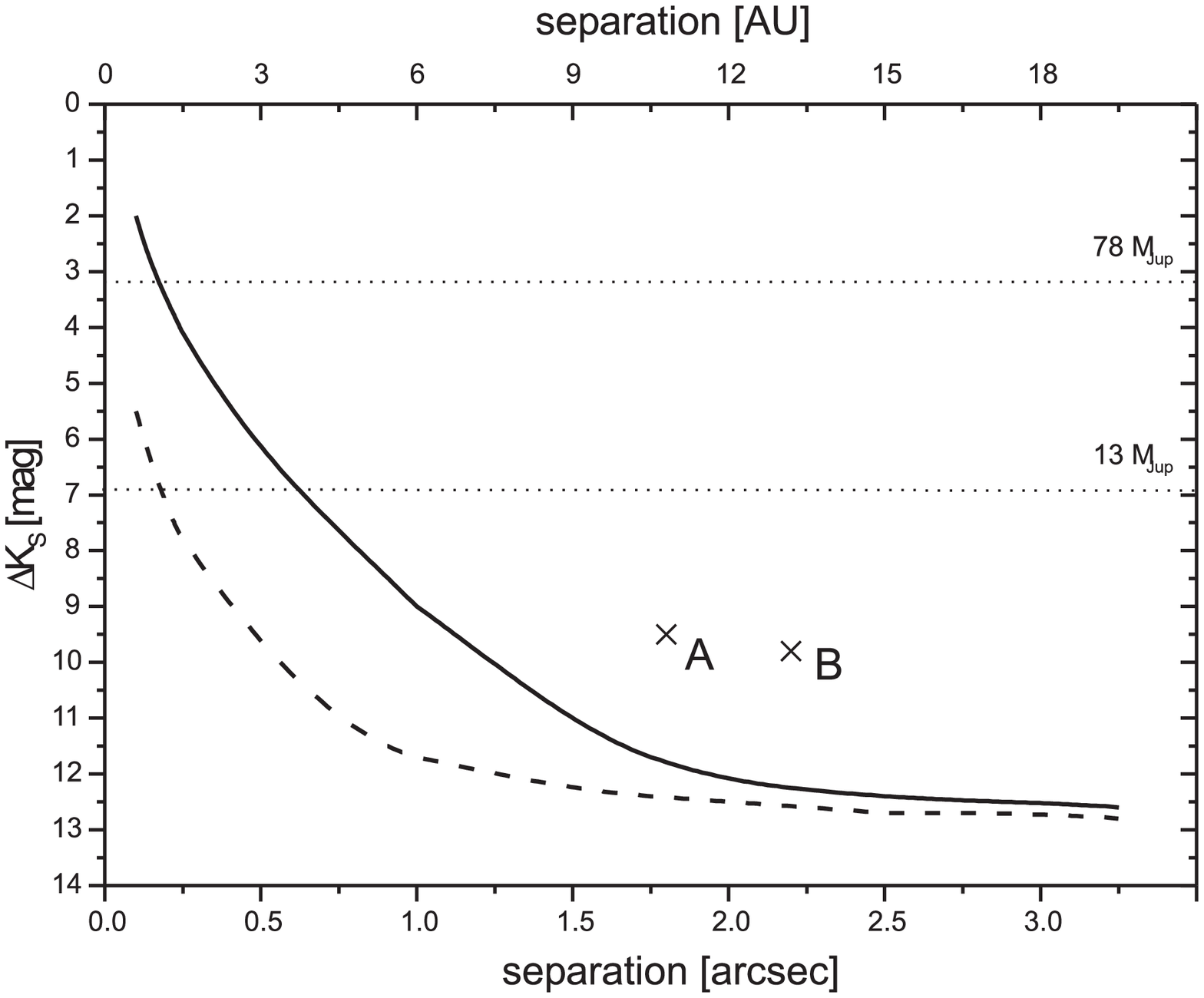}
\includegraphics[height=0.35\textwidth]{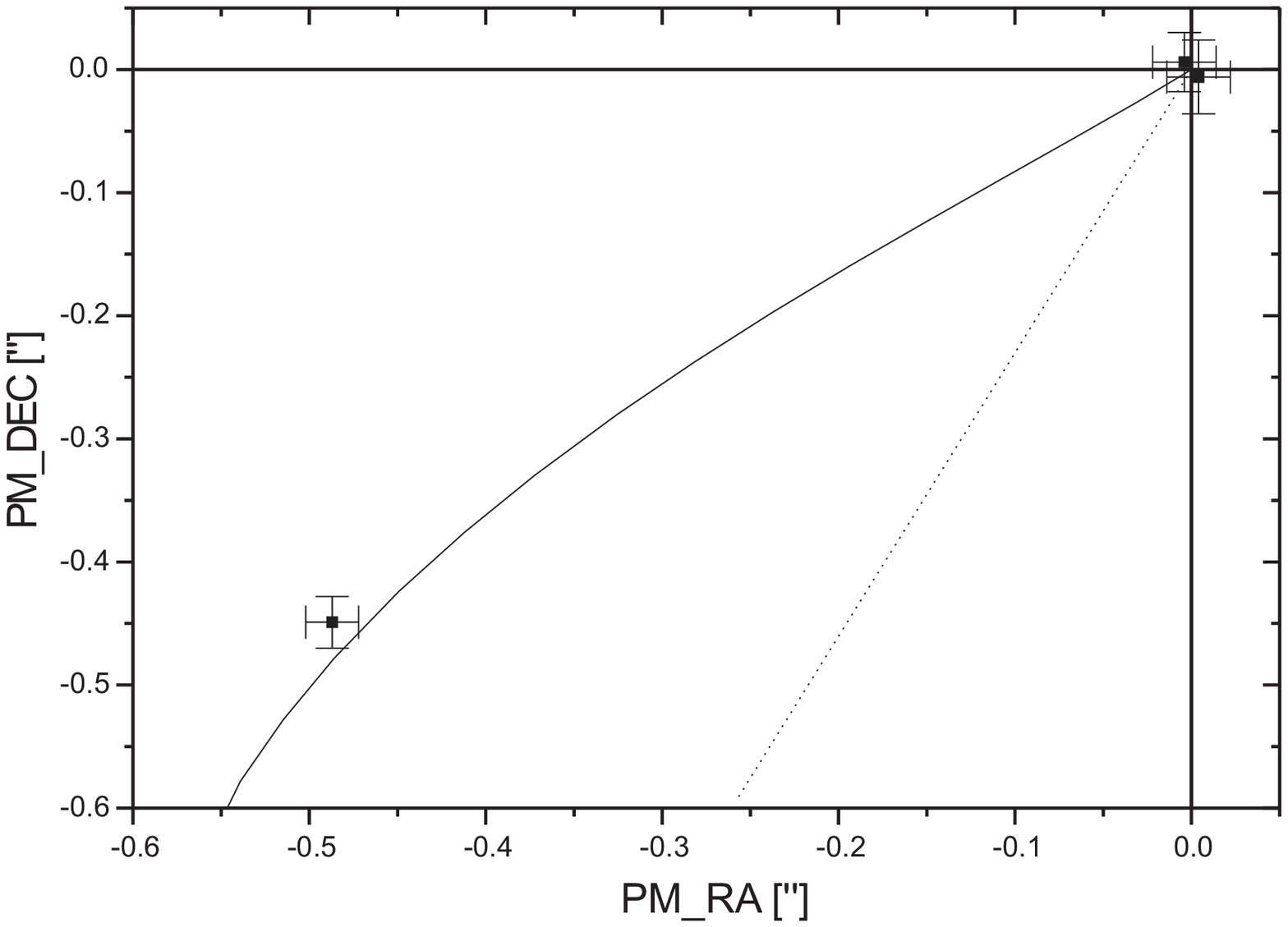}
\end{center}
\caption[]{\textsl{left:} We show here the measured dynamic range for the NACO K$_{S}$ band image
which is shown in Fig.1. The limiting magnitude (S/N=3) for the reduced NACO (solid line) and the
PSF subtracted image (dashed line) is plotted. A\&B indicate two companion candidates.\newline
\textsl{right:} We show the measured astrometric motion of HD\,180617 (black data points) which is
consistent with the calculated motion computed with Hipparcos data for proper and parallactic
motion (solid line). The dashed line is the calculated stellar motion when only proper motion is
considered. The differential atmospheric refraction as well as the errors of the Hipparcos data
aren't considered here. The shown astrometric uncertainties are 3\,$\sigma$ error bars. The dots at
(0,0) are detected companion candidates. They turned out to be non-moving background stars.}
\label{eps3}
\end{figure}

%\appendix

%\section*{Appendix}


\begin{thebibliography}{1.}
\bibitem{Neu1}
\addcontentsline{toc}{section}{References} Neuh\"auser et al.: 'Infrared imaging search for
low-mass companions to members of the young nearby Horologium, Tucana, and $\beta$ Pic
associations' \emph{2003, AN, 324, 535}
\bibitem{Bar1}
Baraffe et al.: 'Evolutionary models for cool brown dwarfs and extrasolar giant planets. The case
of HD 209458', \emph{2003, A\&A 402, 701}

\end{thebibliography}
\end{document}